# Analysis of the Effect of Tilted Corner Cube Reflector Arrays on Lunar Laser Ranging

Jin Cao [1,2,3], Rufeng Tang [1], Kai Huang [1,3], Zhulian Li [1], Yongzhang Yang [1], Kai Huang [2], Jintao Li [1] and Yuqiang Li [1,*]

1 Yunnan Observatories, Chinese Academy of Sciences, Kunming 650216, China; caojin@ynao.ac.cn (J.C.);
2 College of Mathematics and Physics, Leshan Normal University, Leshan 614000, China; hk@lsnu.edu.cn
3 University of Chinese Academy of Sciences, Beijing 100049, China
* Correspondence: lyq@ynao.ac.cn

**Abstract:** This paper primarily investigates the effect of the tilt of corner cube reflector (CCR) arrays on lunar laser ranging (LLR). A mathematical model was established to study the random errors caused by the tilt of the CCR arrays. The study found that, ideally, when the laser ranging pulse width is 10 picoseconds or less, it is possible to distinguish from which specific corner cubes within the CCR array each peak in the echo signal originates. Consequently, partial data from the echo can be extracted for signal processing, significantly reducing random errors and improving the single-shot precision of LLR. The distance obtained by extracting part of the echo can be reduced to the center position of the array, thereby providing multiple higher-precision ranging results from each measurement. This not only improves the precision of LLR but also increases the data volume. A simulation experiment based on the 1.2 m laser ranging system at Yunnan Observatories was conducted. By extracting one peak for signal processing, the single-shot precision improved from 32.24 mm to 2.52 mm, validating the theoretical analysis results. Finally, an experimental laser ranging system based on a 53 cm binocular telescope system was established for ground experiments. The experimental results indicated that the echo signal could identify the tilt state of the CCR array. By extracting the peak returned by the central CCR for signal processing, the ranging precision was greatly improved. Through theoretical analyses, simulation experiments, and ground experiments, a solution to reduce the random errors caused by the tilt of the CCR array was provided. This offers an approach to enhance the single-shot precision of future LLR and provides a reference for upgrading ground-based equipment at future laser ranging stations.

Keywords: lunar laser ranging; corner cube reflector array; lunar librations; laser pulse width; ranging precision

# 1. Introduction

Lunar laser ranging (LLR) is a method of measuring the distance between the Earth and the Moon using a fusion of technologies, including laser technology, computer control, single photon detection, and other technologies. Laser pulses are emitted from Earth-based laser ranging stations towards the corner cube reflector (CCR) array placed on the lunar surface. These reflectors then return the laser pulses back to Earth, where they are detected by single-photon detectors. The precise timing of the laser pulse emission and return is recorded using event timers. By calculating the elapsed time between emission and reception, we determine the distance between the Earth and the Moon[1–3].In the early stages of LLR, the ranging precision for a normal point was approximately 30 cm. With improvements to the timer of the 2.7 m LLR system at the McDonald Observatory, the ranging precision at the McDonald Observatory was increased to 15 cm by the mid-1970s[1].Ten years later, the McDonald Observatory used a new 0.76 m telescope and improved lunar ranging precision to 5 cm. In 1990, the Grasse Laser Ranging Station in France began LLR operations, achieving a normal-point precision of 2--3 cm. Subsequently, when the station conducted LLR experiments using a 1064 nm laser, the results showed a significant increase in echo photons compared to those obtained using the 532 nm laser. Furthermore, the issue of being unable to perform LLR during new moon and full moon phases was effectively resolved. Later experiments successfully demonstrated the feasibility of conducting LLR during daytime[4,5].By the end of 2005, the Apache Point Observatory Lunar Laser-ranging Operation (APOLLO) upgraded its equipment with a larger-aperture (3.5 m) telescope and more advanced detection technology (array detectors), achieving ranging precision at the level of a few millimeters[6].With the continuous improvement in ranging precision, LLR has provided more precise measurements for tests of equivalence principles, time variation of the gravitational constant, gravitational magnetic effects, Earth tidal dissipation, and other gravity theories, thus offering higher-precision examinations for general relativity and other gravitational theories[2,3,7,8].Meanwhile, LLR data can be applied to research of the dynamics of the Earth--Moon system, exploration of the Moon's internal structure, astrophysics, and other fields[9,10].A normal point is a combination of many single-shot measurements, achieving millimeter-level precision by averaging thousands of echo photons. However, even for APOLLO, which has relatively high LLR precision, the single-shot range precision is still at the level of a few centimeters. The single-shot ranging precision of the Grasse laser ranging station in France is approximately 5 cm[5].Murphy et al. indicated that for APOLLO, laser pulse broadening is the largest source of random error in LLR, reaching 100 to 300 ps. As the technology of laser ranging stations advances, other random errors have been significantly reduced to approximately 52 ps[11–13].The broadening of the laser pulse is primarily caused by the varying distances to different reflectors in the tilted CCR array. Table1 shows the single-point ranging errors of APOLLO. It can be seen from the table that the largest source of error is the tilt of the lunar reflector array[14].To further advance fundamental science in related fields, higher-precision LLR data are required[15].Researchers worldwide are actively studying ways to improve LLR precision. Martini et al. from Italy and Turyshev et al. from the United States have proposed placing a single large-aperture reflector on the lunar surface to reduce random errors caused by the tilt of the CCR array[11,15,16] ,However, as of now, no country has yet deployed such devices for ground-based LLR on the lunar surface. Dehant et al. proposed the differential lunar laser ranging (DLLR) technique, which can significantly reduce errors caused by laser transmission through the atmosphere[17,18] ,but it does not address the significant random errors caused by the tilt of the CCR array. Samain et al. treated the CCR array as a whole and estimated the random errors induced by the angle between the incident laser and the array panels due to lunar libration[12].This paper focuses on the lunar surface CCR array, modeling each CCR individually to analyze the effect of array tilt on LLR precision. Based on the analysis, we propose recommendations for potentially reducing random errors caused by array tilt and improving LLR precision in the future.

Table 1 **APOLLO RANDOM ERROR BUDGET PER PHOTON**[14]

| Random Error Source | Time Uncertainty (ps) | Range Error (mm) |
|---|---|---|
| Retroreflector Array Orientation | 100-300 | 15-45 |
| APD Illumination | 60 | 9 |
| APD Intrinsic | 50 | 7 |
| Laser Pulse Width | 45 | 6.5 |
| Timing Electronics | 20 | 3 |
| GPS-disciplined Clock | 7 | 1 |
| Total Random Uncertainty | 136-314 | 20-47 |

## 2 Theoretical Analysis

## 2.1 Effect of CCR Array on the Precision of LLR

### 2.1.1 CCR Arrays

In July 1969, Apollo 11 successfully landed on the moon, and astronaut N. Armstrong placed the CCR array at the predetermined location on the lunar surface. Subsequently, the United States also placed CCR arrays on the lunar surface during the Apollo 14 and Apollo 15 missions[2]. In addition, the former Soviet Union's Luna 17 and Luna 21 missions deployed the Lunokhod 1 and 2 lunar rovers, each equipped with CCR arrays manufactured in France, on the lunar surface. Since then, research on LLR has been conducted by several countries, accumulating over 50 years of observational data[19]. The CCR arrays deployed during the Apollo 11 and Apollo 14 missions consist of 100 corner cubes, each with a diameter of 38 mm. These corner cubes are arranged in a square array measuring 460 mm by 460 mm, with ten corner cubes forming a row and a total of ten rows[20]. The CCR array deployed during the Apollo 15 mission consists of 300 corner cubes, each with a diameter of 38 mm. It measures 1050 mm by 640 mm in size. The array comprises two parts: one part consists of 17 rows and 12 columns, totaling 204 corner cubes, while the other part consists of 8 rows and 12 columns, totaling 96 corner cubes[21]. The CCR array placed on the Soviet lunar rovers consists of 14 equilaterally triangular prisms, each with a side length of 106 mm. These prisms are evenly arranged in two rows[11]. The corner cube prisms that make up the CCR array are tetrahedral structures with three mutually orthogonal reflective surfaces meeting at a common vertex. When a light beam enters the prism through its top surface, it undergoes three sequential reflections on the orthogonal surfaces and returns along the incoming direction. The existing reflectors on the lunar surface are all arrays composed of multiple CCRs. These arrays are designed to increase the reflective surface area and enhance the photon return rate.The photos of the five CCR arrays placed on the lunar surface are shown in Figure 1. Apollo 15, with its largest area, provides the most LLR data. The distribution of the five existing CCR arrays on the lunar surface is illustrated in Figure 2. When placed on the lunar surface, they exhibit certain deviations from directly facing the Earth's center. The pointing deviation of the Apollo-series CCR arrays is approximately ±1°, while that of the Lunokhod CCR arrays is even larger, around ±5°[11,12].

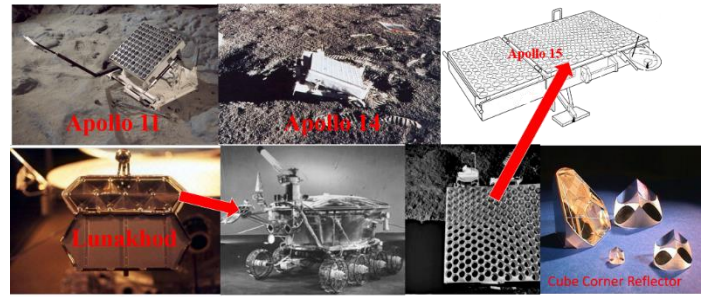

Figure 1 CCR arrays. The figure displays all the CCR arrays on the lunar surface, along with some additional details. (Source: adapted from an image search result for "lunar corner cube reflector" on Bing, https://bing.com/, accessed on 5 May 2024).

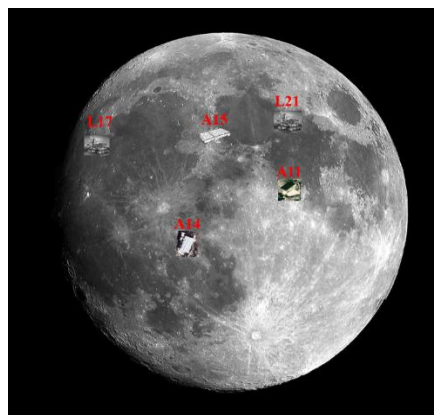

Figure 2 The CCR array sites on the Moon. (Source: adapted from an image search result for "lunar corner cube reflector" on Bing, https://bing.com/, accessed on 5 May 2024).

### 2.1.2 Lunar Libration

Due to the gravitational forces, inertia, and historical evolution of the Earth--Moon system, the Moon's rotation period is very close to its orbital period around the Earth, causing the same side of the Moon to always face the Earth[22,23]. Lunar libration describes the oscillation of the Moon's motion in space, which is a combined result of the Moon's rotation, its orbital path, and the gravitational forces exerted by surrounding celestial bodies[24,25]. Libration is categorized into geometric libration and physical libration. Due to the uneven distribution of the Moon's mass, its rotation is irregular, causing a deviation from uniform rotation and resulting in a lack of complete synchronization between the Moon's rotation and orbital motion. This phenomenon leads to what is known as geometric libration. The Moon's physical libration is categorized into forced libration and free libration. Forced libration is caused by time-varying torques on the Moon's shape due to the gravitational influence of the Sun, Earth, and other planets. Free libration is primarily caused by geological activity, such as intense impacts, interactions between the Moon's core and mantle, and resonance with forced libration. In the latest research, scientists used the high-precision lunar ephemeris DE430 to identify three physical quantities describing the Moon's physical libration, with amplitudes of approximately 200 arcseconds, which is much smaller than the amplitude of the Moon's geometric libration. Therefore, the libration effect observed from Earth is mainly caused by the Moon's geometric libration[26]. Based on the data from the lunar ephemeris, we plotted the line graph of the lunar libration amplitude over 10 years (Figure 3) and over 1 year (Figure 4). From the graphs, it can be observed that the lunar libration amplitude in the latitude direction can reach up to ±6.8°, while in the longitude direction, it can reach up to ±8°. The CCR arrays placed on the lunar surface vibrate with lunar libration, causing them to tilt relative to the direction of laser incidence. Table 2 presents a statistical analysis of the full-rate data from the Geodetic Observatory Wettzell. It calculates the mean lunar libration values corresponding to the station's position for each ranging time interval as well as the echo data width and root mean square (RMS). From the table, it can be observed that the tilt angle in the latitude direction remains relatively consistent. As the tilt angle increases in the longitude direction, both the echo data width and the RMS of the ranging data gradually increase. This indicates that with the influence of lunar libration, larger tilt angles result in larger errors in LLR and lower data precision.

**Table 2** Partial measured data table for the Geodetic Observatory Wettzell (49.1444◦N, 12.8780◦E). The data statistically analyzed in the table are all sourced from the International Laser Ranging Service (ILRS),Available online: https://ilrs.gsfc.nasa.gov, accessed on 5 January 2024.

| Data | Time/UTC | Lat/° | Lon/° | Data Width/ps | RMS/mm |
| --- | --- | --- | --- | --- | --- |
| 20221017 | 040747-074936 | -6.20 | -0.35 | ±150ps | 19.64 |
| 20230302 | 185320-221847 | -5.97 | 2.93 | ±200ps | 24.00 |
| 20230301 | 172841-173459 | -6.05 | 3.40 | ±300ps | 33.33 |
| 20220314 | 195433-214837 | -6.42 | -4.95 | ±400ps | 57.61 |
| 20220411 | 173054-212050 | -6.71 | -5.78 | ±400ps | 59.94 |

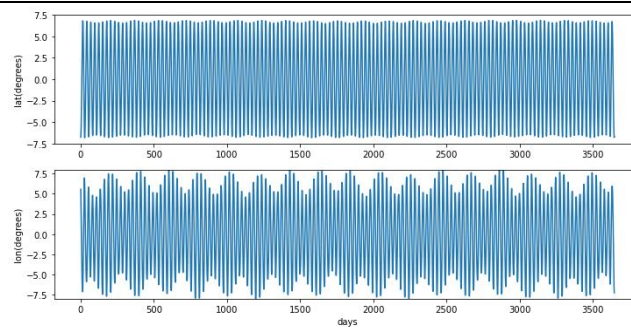

Figure 3 Lunar libration amplitude (1 January 2000–31 December 2009. 10 years).

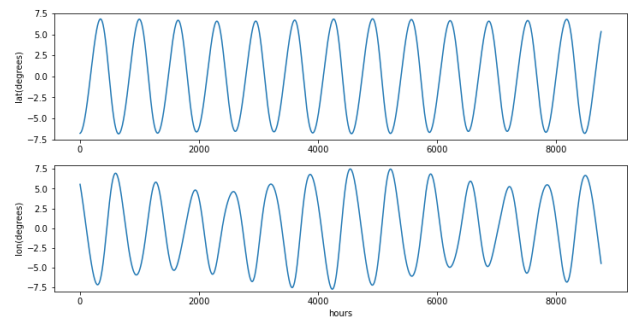

Figure 4 Lunar libration amplitude (1 January 2000–31 December 2000. 1 year).

### 2.1.3 Laser Broadening

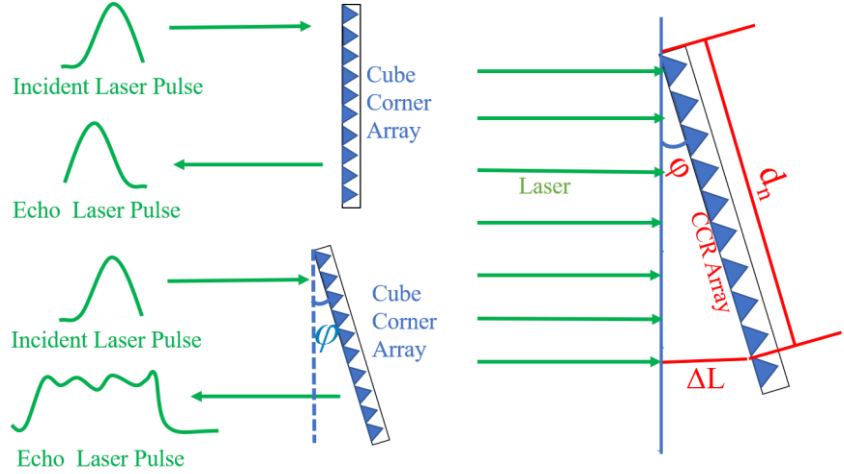

Figure 5 Schematic diagram of a tilted CCR array with laser incidence.

Due to the initial pointing error when placing the CCR array on the lunar surface and lunar libration, the array tilts relative to the direction of laser incidence. As shown in Figure 5, the tilted CCR array causes laser pulses to reach different CCRs at varying distances. When a laser pulse is incident on a CCR array with an inclination angle of φ, the distance difference between the laser pulse traveling to the nearer and farther CCRs is ΔL：

$$\Delta L = d_n * \sin(\varphi) \tag{1}$$

In the equation, $d_n$ represents the central spacing between the nearer and farther CCRs of the CCR array. Taking Apollo 11 as an example, the spacing between adjacent CCRs is 46 mm. We can calculate ΔL and the time difference Δt required for the round-trip flight of the laser pulse, as shown in Table 3. When the tilt angle is 8°, the difference in the laser pulse travel distance to the centers of two adjacent corner cube reflectors is 6.402 mm. From this, it can be deduced that the distance difference between the centers of the farthest-separated two rows of CCRs is 57.6 mm. The round-trip flight time for the laser pulse is 384.4 ps. Therefore, after the laser pulse is reflected by CCRs at different distances, the reflected pulses overlap. The overlapped laser pulse is broadened, increasing the random error in the detection of echo signals. To reduce the random errors caused by the tilt of the CCR array and improve LLR precision, the next step involves studying the CCR array. A mathematical model will be developed to quantitatively analyze the broadening of laser echoes due to CCR array tilt. Subsequently, simulations will be conducted to explore echo detection.

Table 3 Table of distance differences and round-trip flight time differences of laser pulses to adjacent reflectors at different tilt angles.

| Tilt Angle/° | ΔL /mm | Δt /ps |
|---|---|---|
| 2 | 1.6054 | 10.7 |
| 4 | 3.2088 | 21.4 |
| 6 | 4.8083 | 32.1 |
| 8 | 6.4020 | 42.7 |

## 2.2 Mathematical Model

### 2.2.1 CCR Array Model

Using the Apollo 11 and 14 CCR arrays as examples (the structures of the Apollo 11 and 14 CCR arrays are identical), a mathematical model is established to calculate the laser echo pulse envelope. A Cartesian coordinate system is set up with the center of the CCR array (denoted as O) as the origin. As shown in Figure 6, the XOY plane corresponds to the plane in which the array is positioned when it faces directly towards the Earth. Here, the X-axis represents the direction along the latitude lines of the array, and the Y-axis represents the direction along the longitude lines of the array. The Z-axis points vertically towards the center of the Earth, perpendicular to the XOY plane.

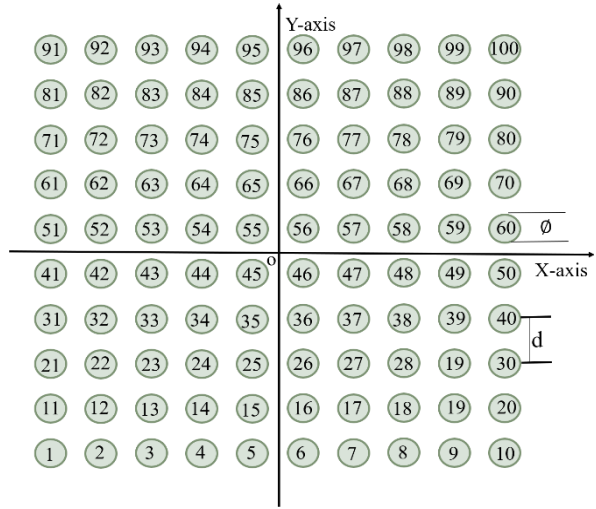

Figure 6 Schematic diagram of the Apollo 11 and 14 LLR reflector arrays (d = 46 mm, φ = 38 mm).

We establish an **R** matrix to represent the coordinates of all individual CCRs. Since the array is in the XOY plane, $z_1, \ldots, z_{100}$ are all 0.

$$\mathbf{R} = \begin{pmatrix} x_1 & \cdots & x_{100} \\ y_1 & \cdots & y_{100} \\ z_1 & \cdots & z_{100} \end{pmatrix} \quad (2)$$

Due to the effect of lunar libration, the deviation angle of the array in the latitude direction is α, meaning that the array rotates around the X-axis by an angle α. The **R** matrix thus becomes **R′**:

$$\mathbf{R'} = \begin{pmatrix} 1 & 0 & 0 \\ 0 & \cos\alpha & -\sin\alpha \\ 0 & \sin\alpha & \cos\alpha \end{pmatrix} \mathbf{R} \quad (3)$$

The array also undergoes a deviation in the longitude direction, with a deviation angle of β, meaning the array rotates about the Y-axis by an angle of β. The coordinate matrix of the CCRs undergoes another transformation, denoted as **R″**:

$$\mathbf{R}'' = \begin{pmatrix} \cos\beta & 0 & -\sin\beta \\ 0 & 1 & 0 \\ \sin\beta & 0 & \cos\beta \end{pmatrix} \mathbf{R}' \quad (4)$$

The angle φ between the direction vector **L**= (0, 0, -1), of the laser incident direction relative to the coordinate system established by the CCR array and the normal vector **F**= (cosαsinβ, sinα, cosαcosβ) of the CCR array after rotation represents the laser incident angle. The projection of the LLR CCR array onto the laser incident direction is:

$$\mathbf{P} = \mathbf{L}\mathbf{R}'' \quad (5)$$

Therefore, the vector representing the time difference $\Delta t_i$ for the laser to reach each CCR relative to point O is denoted as **T**:

$$\mathbf{T} = \frac{\mathbf{P}}{c} \quad (6)$$

where $\Delta t_1$, $\Delta t_2$ …… $\Delta t_n$ are its elements, and c is the speed of light in vacuum. Calculating the RMS of the time series **T** yields the laser pulse broadening caused by the tilt of the CCR array (as shown in Table 4). For Apollo 15, establishing the same model produces the following results (as shown in Table 5), which are generally consistent with the results published by APOLLO[27].

Table 4 The dispersion in picoseconds (RMS) varies as a function of the longitude and latitude angles (in degrees) for the Apollo 11 and 14 LLR reflector arrays.

| LON/° \ LAT/° | 0 | 2 | 4 | 6 | 8 |
|---|---|---|---|---|---|
| 0 | 0.0 | 30.8 | 61.5 | 92.1 | 122.7 |
| 2 | 30.8 | 43.5 | 68.7 | 97.1 | 126.4 |
| 4 | 61.5 | 68.7 | 86.8 | 110.6 | 137.0 |
| 6 | 92.1 | 97.1 | 110.6 | 129.9 | 152.9 |
| 8 | 122.7 | 126.4 | 137.0 | 152.9 | 172.6 |

Table 5 The dispersion in picoseconds (RMS) varies as a function of the longitude and latitude angles (in degrees) for the Apollo 15 LLR reflector array.

| LON/° \ LAT/° | 0 | 2 | 4 | 6 | 8 |
|---|---|---|---|---|---|
| 0 | 0.0 | 72.0 | 143.8 | 215.5 | 287.0 |
| 2 | 39.4 | 82.0 | 149.0 | 219.0 | 289.5 |
| 4 | 78.8 | 106.6 | 163.8 | 229.2 | 297.2 |
| 6 | 118.1 | 138.1 | 185.7 | 245.2 | 309.6 |
| 8 | 157.3 | 172.7 | 212.6 | 266.0 | 326.2 |

The laser pulse waveform emitted by the laser emitter follows a Gaussian distribution, expressed as :

$$s(t) = \frac{1}{\sqrt{2\pi}\tau_s} \exp\left(-\frac{t^2}{2\tau_s^2}\right) \quad (7)$$

The laser pulse waveform reflected by each CCR is:

$$S(t_i) = \frac{1}{\sqrt{2\pi}\tau_s} \exp\left[-\frac{(t-2\Delta t_i)^2}{2\tau_s^2}\right] \quad (8)$$

Where $\tau_s$ is related to the laser's full width at half maximum (FWHM); $\tau_s$ = FWHM/ 2.3548. Therefore, the laser waveform reflected by the LLR reflector array is the superposition of the laser waveforms reflected by each CCR. The time-domain distribution function is given by:

$$S(t) = \sum_{i=1}^{n} S(t_i) \tag{9}$$

## 2.2.2 Echo and Noise Energy Model

The detection of space objects relies on estimating the number of return photons in the ranging system based on the laser ranging radar equation. The estimation of return photons in LLR still employs the laser ranging radar equation[28]:

$$N_e = \eta_q \left( E_r \frac{\lambda}{\hbar c} \right) \eta_t G_t \sigma \left( \frac{1}{4\pi D^2} \right)^2 A_r \eta_r T_a^2 T_c^2 \tag{10}$$

In the equation, $N_e$ represents the mean number of photoelectrons recorded by the detector; $\eta_q$ is the detection efficiency of the detector; $\lambda$ is the wavelength of the ranging laser; $h$ is the Planck constant; $c$ is the speed of light; $\eta_t$ is the transmit optics efficiency; $G_t$ is the the transmitter gain; $\sigma$ is the target optical cross-section; $D$ is the predicted distance from the ranging station to the moon; $A_r$ is the effective area of the telescope receive aperture; $\eta_r$ is the efficiency of the receive optics; $T_a$ is the one-way atmospheric transmission; $T_c$ is the one-way transmissivity of cirrus clouds.

$$D = -(R_E + h_t)\cos\theta_{zen} + \sqrt{(R_E + h_t)_2 \cos^2\theta_{zen} + 2R_E(h_s - h_t) + h_s^2 - h_t^2} \tag{11}$$

where $R_E$ is the radius of the Earth, $h_t$ is the altitude of the ranging station, $h_s$ is the altitude of the Moon relative to sea level, and $\theta_{zen}$ is the zenith angle of the Moon observed from the station.

$$G_t(\theta) = \frac{8}{\theta_t^2} \exp\left[-2\left(\frac{\theta}{\theta_t}\right)^2\right] \tag{12}$$

In the above formula, $\theta_t$ is the far-field divergence angle of the beam, and $\theta$ is the pointing error of the beam. For a space target equipped with a CCR, the target optical cross-section $\sigma$ is:

$$\sigma = \rho A_{ccr} \left( \frac{4\pi}{\Omega} \right) \tag{13}$$

where $\rho$ is the reflectivity of the CCR, and $\Omega$ is the reflection divergence angle of the CCR. For a CCR with a radius of $R_{ccr}$, when the incident angle of the laser is $\varphi_{in}$, the effective reflective area Accr of the CCR is:

$$\begin{cases} A_{ccr} = n_{ccr}\pi R_{ccr}^2 \eta(\varphi_{in}) \\ \eta(\varphi_{in}) = \dfrac{2}{\pi}\left(\sin^{-1}\mu - \sqrt{2}\mu\tan\varphi_{ref}\right)\cos\varphi_{in} \\ \varphi_{ref} = \sin^{-1}\left(\dfrac{\sin\varphi_{in}}{n_{ref}}\right) \\ \mu = \sqrt{1 - 2\tan^2\varphi_{ref}} \end{cases} \quad (14)$$

where $n_{ref}$ is the refractive index of the CCR material (with a value of 1.455), and $n_{ccr}$ is the number of CCRs. When the laser is incident perpendicularly, $\eta(\varphi_{in})$ equals 1.

The system simulation is based on the 1.2 m telescope system at Yunnan Observatories as a reference. The parameters of the 1.2 m telescope laser ranging system are shown in Table 6.

Table 6 Table of parameters for the 1.2 m LLR system at Yunnan Observatories.

| Parameter | Value |
| --- | --- |
| Laser wavelength ($\lambda$) | 532 nm |
| Pulse energy ($E_r$) | 0.1 J |
| Laser repetition rate | 100 Hz |
| Telescope aperture | 1.06 m |
| Laser divergence angle ($\theta_t$) | 2″ |
| Pointing error ($\theta$) | 1″ |
| Reflector divergence angle ($\Omega$) | 8″ |
| Detector efficiency ($\eta_q$) | 0.2 |
| Transmit optics efficiency ($\eta_t$) | 0.4 |
| Receive system efficiency ($\eta_r$) | 0.2 |
| Atmospheric transmittance ($T_a$) | 0.6 |
| Transmissivity of cirrus clouds ($T_c$) | 1 |
| Average echo photons for A11 ($N_e$) | 0.079 |

Therefore, the temporal function of the photocounts of the echo pulse can be represented as:

$$G(t) = N_e \cdot S(t) \quad (15)$$

For single-photon detectors, the detection noise mainly arises from background noise and detector dark counts, which are generally assumed to follow a uniform distribution. The average number of photocounts generated by noise during any time interval is typically given by:

$$V_{noise} = N_b \cdot \eta_t + N_{darkcount} \quad (16)$$

where $N_b$ represents the number of background noise photons per unit time, and $N_{darkcount}$ represents the noise from detector dark counts, which is the average number of photocounts generated by the detector itself per unit time.

2.2.3 Detection Model

Yunnan Observatories' LLR system employs a single-photon detector. The number of photoelectrons generated by the conversion of laser echo signals follows a Poisson distribution. Specifically, the probability of generating $k$ photoelectrons when the average number of photoelectrons per single pulse is $N$ can be expressed as:

$$P(k;N) = e^{-k}\frac{N^k}{k!} \tag{17}$$

In the absence of noise consideration, at least one photoelectron needs to be generated to trigger the detector. Therefore, the detection probability is given by:

$$P_t = 1 - P(0, N_e) = 1 - e^{-N_e} \tag{18}$$

In the equation, $P(0, N_e)$ represents the probability of detecting zero photons. The assumption is that the total noise rate during the detection process is $V_{noise}$, the pulse width of the echo signal is $T_{pulse}$, and the echo signal appears at a position $T_{gate}$ after the gate opens. Therefore, the premise for the detector to be triggered by the signal is that within the time of $T_{gate} + T_{pulse}$ after the gate opens, the detector is not triggered by noise. The probability that the detector is not triggered by noise within $T_{gate}$ time is given by::

$$P_n = P(0; N_{noise}) = e^{-N_{noise}} \tag{19}$$

The probability of successful signal detection, considering $N_{noise} = (T_{gate} + T_{pulse}) \cdot V_{noise}$ as the noise count, can be expressed as follows:

$$P_d = P_n \cdot P_t \tag{20}$$

The detection probability in any time interval from $t_1$ to $t_2$ can be expressed as the probability of detecting within the interval from the opening time 0 to $t_2$ while not detecting within the interval from 0 to $t_1$, given by:

$$P(t_1, t_2) = \exp\left[-\int_0^{t_1} f(t)dt\right] - \exp\left[-\int_0^{t_2} f(t)dt\right] \tag{21}$$

The times $t_1$ and $t_2$ here should be any two moments within the detector's response time interval, and it is required that $t_1 < t_2$.

$$f(t) = G(t) + N_{noise} \tag{22}$$

The ranging residuals are obtained by processing the detected photoelectron signals against the predicted values. Signal identification, Poisson statistical filtering, polynomial fitting filtering, etc., are performed on the residual data to obtain the fitted residuals. Data with fitted residual values greater than 2.5 times the standard deviation are removed, and the standard deviation of the remaining data is calculated (internal fitting precision). If the precision does not meet the requirements, the fitting order is increased until the precision meets the requirements. The standard deviation of the final fitted residuals represents the internal fitting precision of the data, which is the single-shot ranging precision.

## 2.3 Results Analysis

Based on the above theoretical model, numerical simulations were conducted, and a simulation system was established using Yunnan Observatories' 1.2 m LLR system. The echo envelopes of laser pulses with different pulse widths incident on CCR arrays with different tilt angles were calculated and analyzed. Simulations of laser ranging with various pulse widths were also performed.

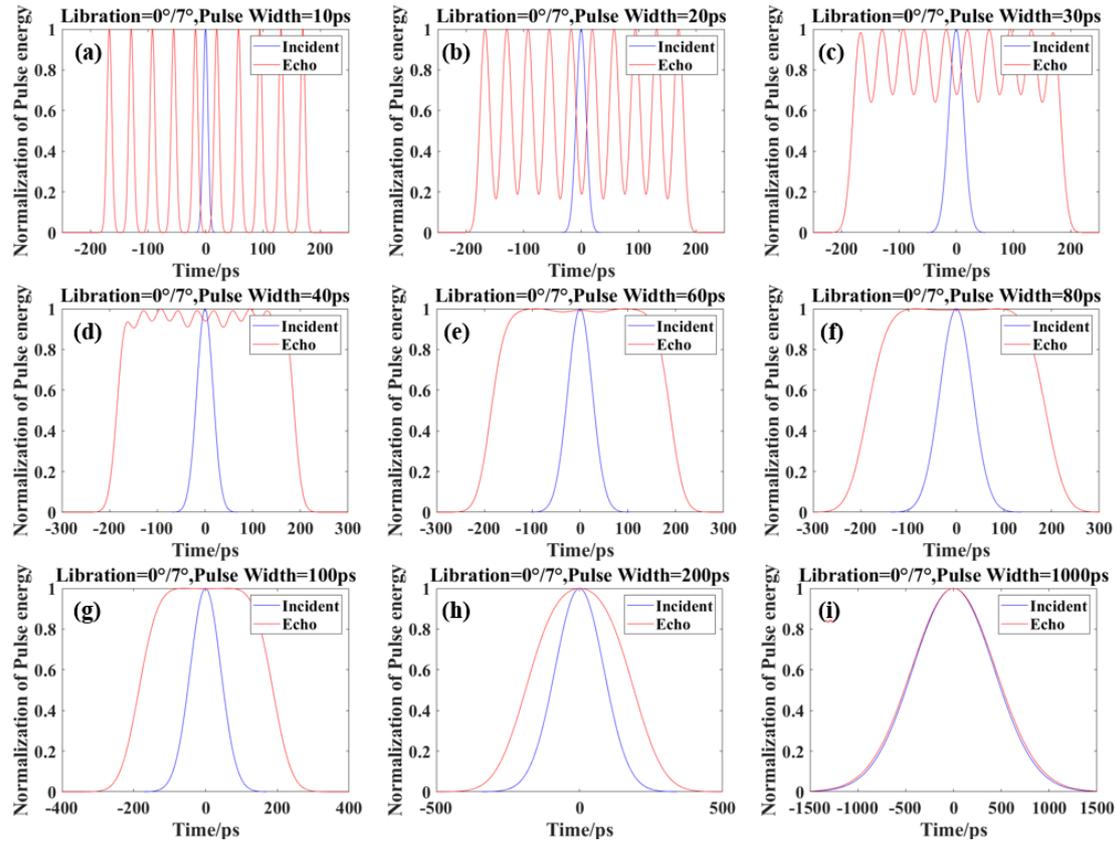

Figure 7 Envelope of different laser pulses due to the tilt of the CCR array (Apollo 11 and 14)

Figure 7 illustrates the laser echo envelopes generated by laser incidences with different pulse widths on the Apollo 11 and 14 CCR arrays, where the arrays are inclined at 0° in latitude and 7° in longitude. The analysis indicates that the tilt of the CCR arrays results in varying times for the laser pulse to reach each CCR. This causes the reflected laser pulses to overlap, leading to pulse broadening. The extent of echo broadening varies with the width of the incident laser pulses: the narrower the incident laser pulse, the more pronounced the echo broadening. When the incident laser pulse width is 1 ns, the width of the echo signal matches the incident width, and the pulse broadening due to the tilt of the CCRs array is not evident. As the incident laser pulse width decreases, the echo envelope shows multiple peak separations. When the incident laser pulse width is 10 ps, the pulses reflected from the farther CCRs do not have time to overlap with the pulses reflected from the closer CCRs, resulting in distinct peaks in the echo waveform.

Since the Apollo 11 and 14 arrays consist of 10 x10 CCRs arranged in a square, each column has the same distance when tilted only in the longitude direction, resulting in 10 different distance values and hence 10 distinct peaks in the echo waveform. Thus, if it is possible to determine which specific CCRs each peak comes from, independent echo peaks can be intercepted for signal processing. This would significantly reduce the random error in LLR caused by the tilt of the reflector array.

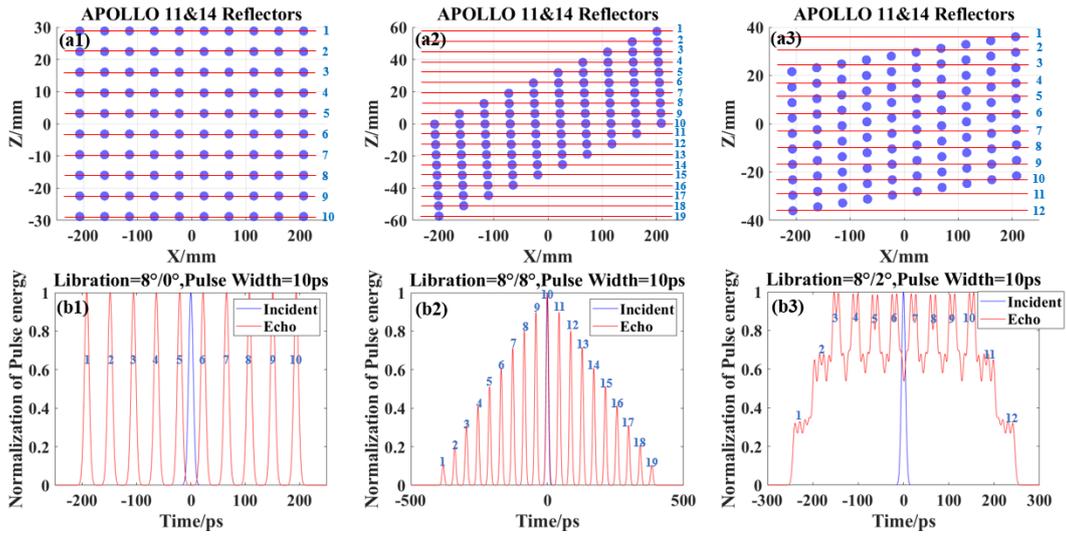

Figure 8 Range echo envelopes of the CCR arrays (Apollo 11 and 14) at different tilt angles.

As shown in Figure 8, a1−a3 represent the tilt states of the LLR CCR arrays, where different CCRs have varying distances to the station along the laser incident direction (z−axis). Reflectors at equal distances are connected by a red line in the figure. Subfigures b1−b3 show the range echo envelopes of 10 ps incident laser pulses reflected by the Apollo 11 and 14 reflector arrays at different tilt states. As shown in the figure above, under ideal conditions, when the incident laser pulse width is 10 ps, the tilt attitude of the CCR array can be inferred from the number of peaks and the shape of the peaks in the laser echo envelope. We can determine which equidistant CCRs the echo signal of each peak originates from.

Based on the analysis of Figures 7 and 8, when the incident laser pulse for LLR is 10 ps or less, a single peak from the echo can be extracted for signal processing. This will significantly reduce the random error caused by the tilt of the CCR array and improve the single−shot precision of LLR. By determining the tilt orientation and angle of the CCR array, the ranging information from each peak can be converted to the distance to the center of the array. In a single laser ranging measurement, this is equivalent to simultaneously ranging multiple different targets, resulting in more comprehensive laser ranging data. Long−term acquisition of such LLR data can be used to infer changes in the tilt state of the CCR array, thereby validating the theory of lunar libration. This provides higher−precision data for the study of Earth−Moon models. Next, the LLR simulation system is used to verify the results of the theoretical analysis by calculating the laser ranging precision when capturing a single peak as the echo signal. The simulation system mainly considers the effect of primary wave detection, echo detection, and the timing system on single-shot ranging precision, with a particular focus on the effect of laser beam broadening due to the tilt of the CCR on the precision of single−shot ranging. The ranging system detects the primary wave emission time using a photodiode, which introduces a random error of less than 5 ps during the laser pulse photon's interaction with the photodiode. In LLR, when the photons from the returning pulse reach the photodetector's sensitive surface, there is a certain randomness in terms of both timing and spatial position, leading to random errors that generally range from 10 to 50 ps. The event timer itself also contributes a random error within 5 ps during operation. To highlight the effect of random errors caused by the tilt of the CCR array, the simulation system sets the random error of primary wave detection at 5 ps, the random error of echo detection at 10 ps, and the random error from the timer and other factors at 5 ps.

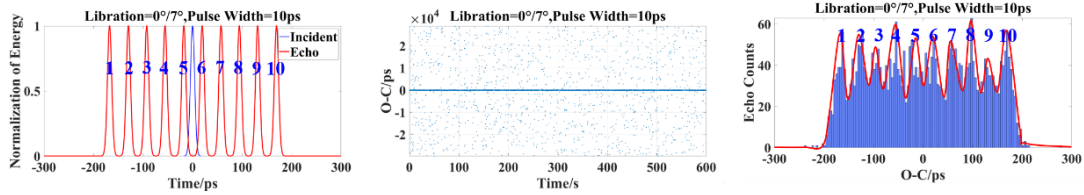

Figure 9 Echo plot for the laser ranging simulation of the Apollo 11 CCR array

Figure 9 shows the theoretically calculated pulse echo envelope, simulated ranging residuals, and histogram of residuals for a 10 ps laser ranging system when the Apollo 11 CCR array is not tilted in the latitude direction and is tilted by 7° in the longitude direction. The parameters used in the simulation system are listed in Table 6. The ranging duration is set to 10 min, with an echo rate of 5.97%. The number of peaks in the echo histogram is 10, with uniform spacing between the peaks. According to the previous theoretical analysis, for the tilted state shown in Figure 8(a1), the average spacing between the peaks is 40.4 ps (12.1 mm). Using Equation (1), the tilt angle is calculated to be 7.5°, which is close to the preset angle. Similarly, the distance from the center of each row of CCRs to the center of the CCR array in the direction of laser incidence can also be calculated. In Figure 9, the single-shot precision of this ranging, calculated using all data points as signals, is 32.24 mm. When considering only the first peak as the ranging signal, the single-shot recision is reduced to 2.52 mm. Each subsequent peak can be treated as an independent echo signal, effectively enabling single-shot ranging for ten targets. Additionally, each ranging result can be adjusted to the center of the CCR array based on the tilt status of the reflectors.

Based on the theoretical analysis and simulation experiments, it can be concluded that when the pulse width of the ranging laser is 10 ps and the energy of a single laser pulse is sufficient to meet the detection requirements for the number of echo photons, it becomes feasible to identify which specific CCRs the different peaks in the echo signal originate from. By capturing a portion of the data for signal processing, the single-shot ranging precision of laser ranging can be improved. To further validate the theoretical results, ground target laser ranging experiments will be conducted. Currently, the laboratory lacks a 10 ps laser suitable for experimentation. According to the calculation method in Table 3, when the spacing between CCRs in the reflector array increases or the tilt angle increases, the time difference for the laser pulse to travel to adjacent CCRs and return becomes greater than the width of the incident laser pulse. This results in the formation of independent peaks in the echo, reflecting the tilted state of the CCR array. Therefore, the next step will involve conducting ground target verification experiments by increasing the spacing between reflectors and increasing the tilt angle of the CCR array.

# 3 Ground Experiment

## 3.1 Setup

The schematic diagram of the ground target experimental system is shown in Figure 10. A 6 × 6 CCR array was constructed based on the structure of the Apollo 11 CCR array. Each individual CCR has a diameter of 25.4 mm, and the center-to-center spacing between CCRs is 46 mm. A manually adjustable two-dimensional tilt table is installed on the back of the CCR array to adjust the tilt angle of the CCRs in both the latitude and longitude directions. The angle adjustment range is ±20◦ (refer to Figure 11). The adjustable CCR array is fixed on the exterior facade of an iron tower wrapped in black velvet cloth, as shown in Figure 12. This iron tower is located at the top of a fire watch tower that is approximately 1.8 km away from the 53 cm binocular telescope at Yunnan Observatories (International Laser Ranging Service station code 7819; site name KUN2; 25.0298◦N, 102.7977◦E). The altitude of this location is slightly higher than that of the 53 cm binocular telescope at Yunnan Observatories.

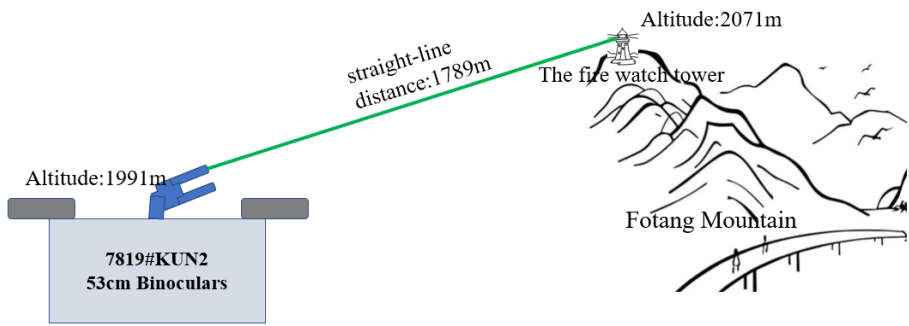

Figure 10 A schematic diagram of the local experiment.

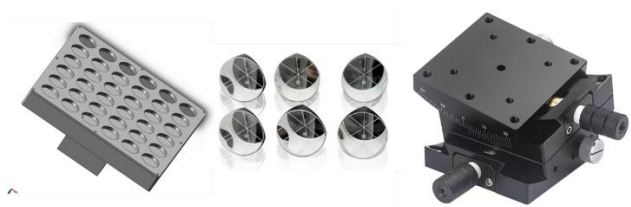

Figure 11 The experimental CCR array (the left image shows the 6 × 6 array of CCRs used in the experiments, the middle image depicts a single CCR, and the right image displays the manually adjustable tilt table that is capable of adjusting angles in two directions).

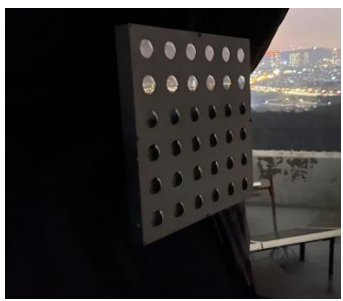

Figure 12 The experimental CCR array fixed on the exterior facade of the iron tower.

The experimental laser ranging system is established on the basis of the Yunnan Observatories' 53 cm binocular telescope, forming a ground target experimental system. The optical system is shown in Figure13, and the main system parameters are listed in Table 7. Before the experiments, personnel on the top of the fire watch tower observe the laser pointing and direct the personnel in the observation room to adjust the telescope's direction so that the laser spot hits the surface of the blank array where no CCRs are installed (as shown in the left image of Figure 14). The laser divergence angle of the laser ranging system is adjusted to ensure the spot covers the entire CCR array. The telescope is kept fixed in this pointing position for the subsequent experiments. A plane mirror is placed at the center of the blank array (as shown in the middle image of Figure 14). A spot inspection plate is used around the plane mirror to observe the laser spot reflected by the mirror. The array's tilt angle is adjusted until the reflected spot is not visible when the spot inspection plate is moved around the mirror's edge, indicating that the laser is perpendicularly incident on the plane of the reflector array. The angles in both directions of the manual tilt stage connecting the CCR array are recorded at this time, and this angle is taken as the initial tilt angle of the array.

Table 7 The main system parameters of the Yunnan Observatories' 53 cm binocular telescope system.

| Parameters | Value |
| --- | --- |
| Aperture(cm) | 51 |
| Laser Wavelength(nm) | 532 |
| Pulse Width (ps) | 100 |
| Repetition Rate(Hz) | 1000 |
| Max. Energy(mJ) | 1 |

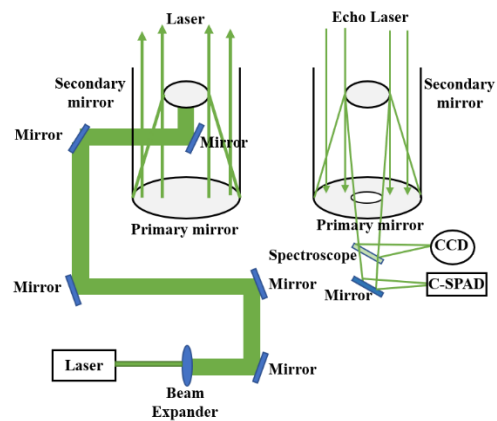

Figure 13 The optical system of the Yunnan Observatories' 53 cm binocular telescope.

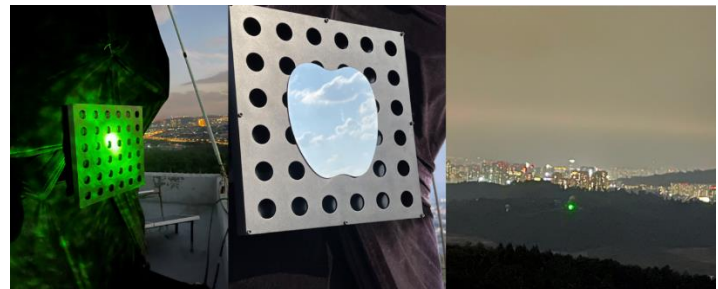

Figure 14 Experimental procedure (left: photo of adjusting the telescope direction and the size of the incident spot; middle: plane mirror attached to the array surface for adjusting the array tilt angle; right: photo of the incident laser.)

Before the experiment, the incident laser energy was attenuated according to the empirical values of the lunar retroreflectors, ensuring that the echo signal reached the single-photon level. The array of CCRs was tilted within the range of ±20° in both the longitude and latitude directions, with a step size of 5°, and the ranging time for each angle was set to 120 s. The main ranging experiments conducted were as follows:

  •Experimenting with laser ranging at various angles by placing a CCR in the center of the blank reflector array;

  • Conducting ranging experiments at various angles with a 36-CCR array;

  • Only the first and sixth columns of the CCR array were kept operational, while the other reflectors were blocked with black shading, and ranging experiments were conducted at various angles;

  • Blocking the second, third, and fifth columns of the array with black shading and conducting

ranging experiments at various angles.

The experimental results presented below show that, except for the experiments with the two-column corner reflectors that were conducted on 25 April 2024, all other experiments were conducted on 27 May 2024.

## 3.2 Experimental Results

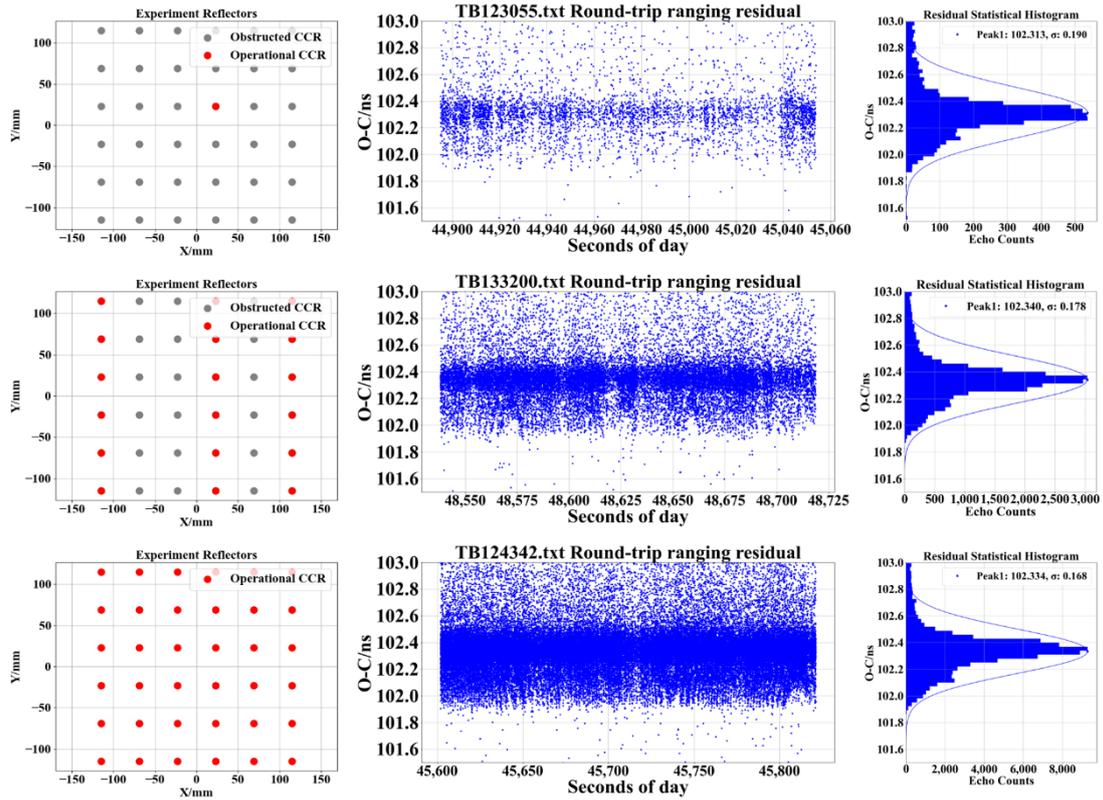

Figure 15 Laser vertically incident echos for different numbers of reflectors.

Figure 15 shows the ranging results for the laser being vertically incident on one CCR, three rows of CCRs, and 36 CCRs. The first column is a schematic diagram of the experimental CCR array, with red indicating operational CCRs and gray indicating obstructed CCRs. The second column is the laser ranging echo residual plot, and the third column is the histogram of the residuals. In the residual plots, ``TB123055'' indicates the experimental data saved at 12:30:55 UTC. Other numbers starting with ``TB'' in the residual plots denote the same meaning. From the figure, it can be seen that the width of the echo data is consistent regardless of the number of reflectors in the CCR array, with only the number of echo photons varying. This indicates that the pre-experimental adjustment of the laser's perpendicular incident angle was effective. The histogram of the laser ranging echoes in the figure is fitted with a Gaussian function, and the center coordinates of the fit represent the central position of the data. The center positions remain stable across the three scenarios, indicating good stability of the experimental laser ranging system.

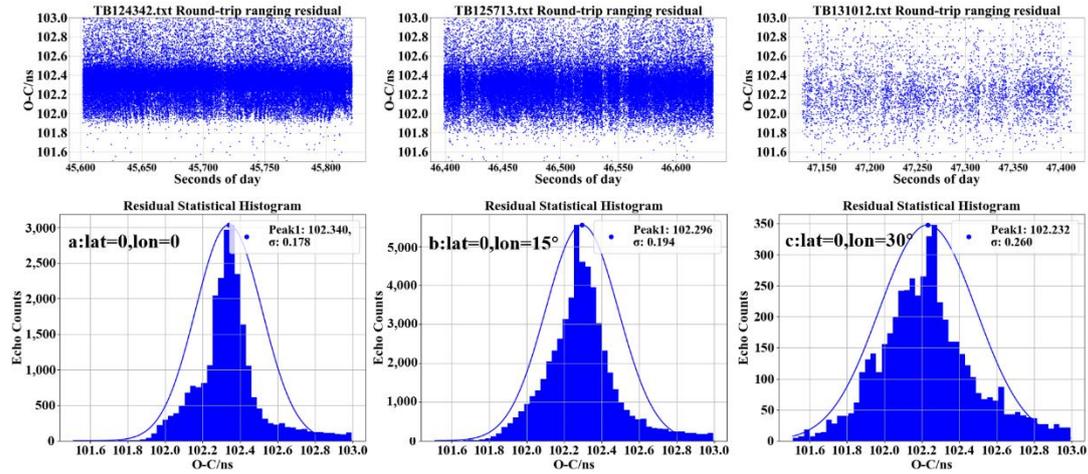

Figure 16 Echoes of the experimental CCR array at different tilt angles.

Figure 16 shows the echo residuals and residual histograms when all 36 CCRs in the experimental array are operational at different tilt angles. In the figure, the CCR array is not tilted in the latitude direction, while the tilt angles in the longitude direction are 0°, 15°, and 30°. As the tilt angle in the longitude direction increases, the sigma of the Gaussian curves gradually increases, indicating a widening of the data width in the echo signals due to the tilt of the CCR array. This widening contributes to an increase in the random error in single-shot ranging. Moreover, as the tilt angle increases, the number of echo photons decreases, which also reduces the precision of the normal points. After processing the ranging results, the single−shot precisions for Figure 16a to Figure 16c are 3.55 cm, 4.50 cm, and 6.92 cm, respectively, fully demonstrating that the tilt of the CCR array significantly affects the precision of laser ranging. This result is consistent with the statistical data obtained from the actual measurements at the Geodetic Observatory Wettzell, as shown in Table 2.

The relationship curve between the peak positions of the echo histograms and the tilt angles of the CCR array under different tilt angles with all 36 CCRs in the experimental CCR array in working condition is plotted in Figure 17. As seen in Figure 17, the peak positions decrease as the tilt angle of the CCR array increases, which leads to a reduction in the calculated measurement distance and affects the accuracy of the ranging results. This phenomenon is caused by the working mechanism of the single-photon detector. Due to the tilt of the reflector array, the distance from individual CCRs to the station varies, making it easier for the photons that return first to trigger the detector. Consequently, the photons reflected by the nearer reflectors are more likely to be detected, causing the entirety of the echo data to shift forward.

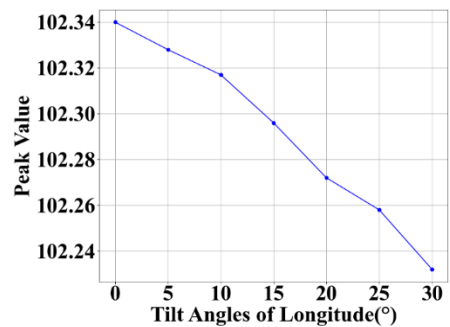

Figure 17 The peak value of the echo histogram changes with the tilt angle of the CCR array.

Figure 18 shows the residuals and histograms of the echo signals for the CCR array at different inclination angles when only the first and sixth columns of CCRs are operational. As the tilt angle increases, the residuals gradually separate into two signal lines, and the histogram of the echo signals gradually splits into two peaks. In figure 18b, the horizontal coordinate difference between the two peaks is 296 ps, corresponding to a distance difference of 44 mm. In figure 18c, the horizontal coordinate difference between the two peaks is 554 ps, corresponding to a distance difference of 83 mm. According to Equation (1), the theoretical distances for CCRs with a tilt angle of 10° and 20° with a column spacing of 230 mm is 40 mm and 79 mm, respectively. The experimental results closely match the theoretical calculations. The few millimeters of difference between the measured and theoretical values may be attributed to significant noise affecting the Gaussian fitting of the signal peaks. Thus, the tilt state of the CCR array can be determined based on the waveform of the ranging echo, and each peak can be attributed to specific CCRs at consistent distances.

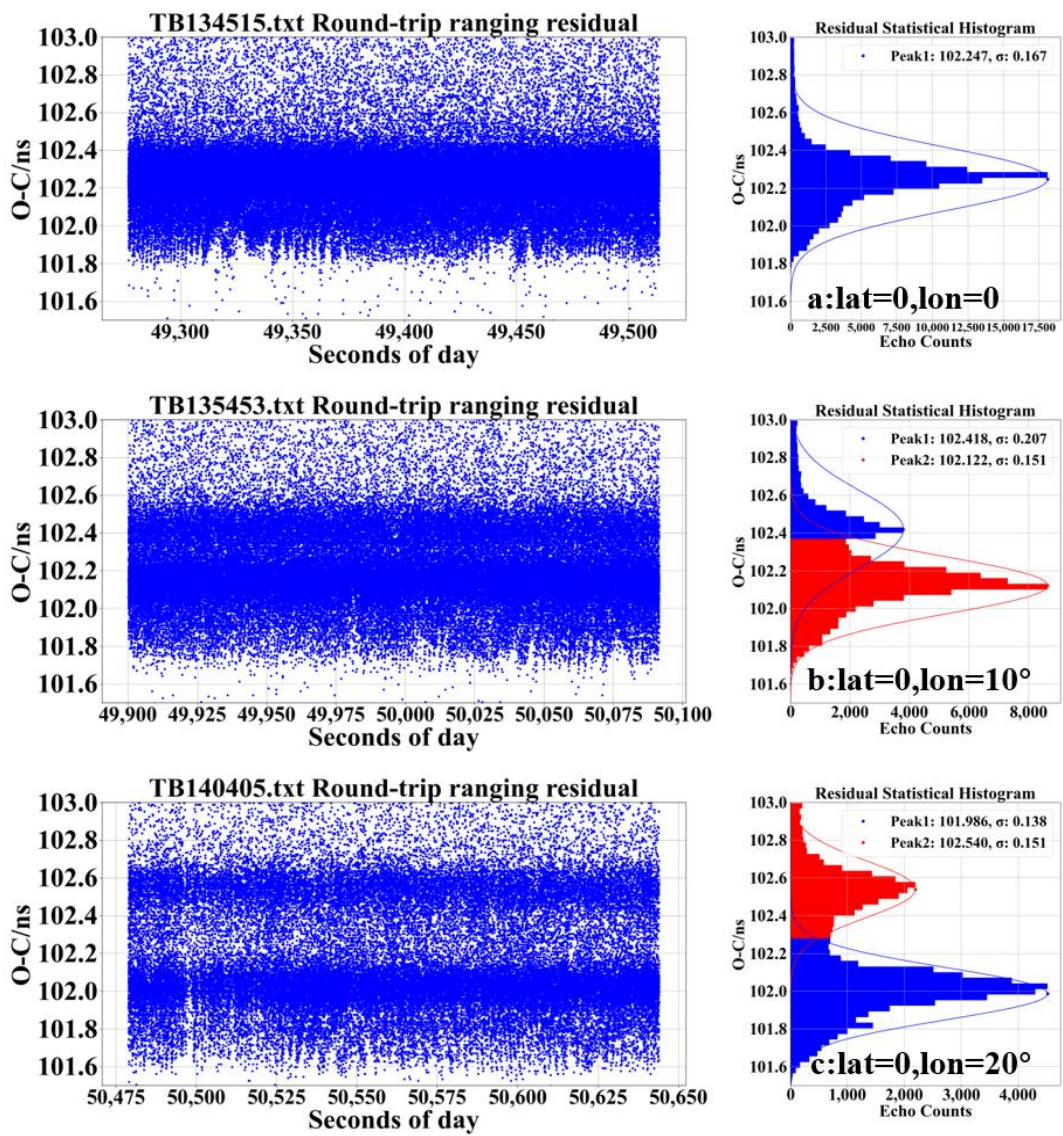

Figure 18 The residual echoes and their histograms at different tilt angles for the experimental CCR array using two columns.

Figure 19 shows the experimental results for a three-column CCR array with a tilt angle of 0° in the latitudinal direction and 30° in the longitudinal direction. The spacing between the blue and green peaks is 474 ps, indicating that the center spacing between the first and second columns of reflectors in the laser incident direction is 71 mm. This corresponds to a laser incident tilt angle of 30.9°, which is consistent with the preset experimental conditions.

Based on the experimental results, the tilt state of the CCR array can be determined. By using the middle peak for signal processing, the single-shot ranging precision is 2.8 cm. Using the entire echo for signal processing, the single-shot ranging precision is 9.6 cm. In Figure 18, the three peaks correspond to echoes reflected by the three columns of CCRs. The echo intensities should be the same, considering the detector's response mode. However, the laser spot was strongest at the center of the CCR array (see Figure 14 left), causing the echo energy from the second column of reflectors to be much greater than that from the other two columns. Thus, the second peak in the measurement data is higher than the first peak, which also reduces the detection probability of the echo from the third column of CCRs, making the third peak very weak.

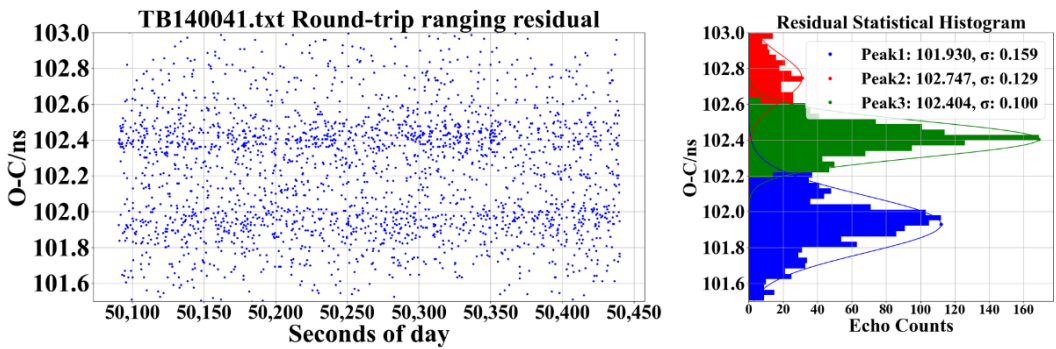

Figure 19 The experimental results for the CCR array with three columns.

## 4 Results

This paper presents a theoretical analysis of the effect of the inclination of lunar CCR arrays on the precision of LLR. The broadening effect of laser pulses due to various tilt angles of existing CCR arrays is calculated, and the envelope of laser ranging echoes is plotted. Based on the analysis of the echo envelope, it is concluded that with a laser pulse width of 10 ps, the tilt state of the CCR array can be identified by analyzing the echo signals, and each peak of the echo can be attributed to a set of equidistant CCRs. Therefore, by intercepting signals reflected from a known set of equidistant CCRs, the random error caused by the tilt of the CCRs can be reduced, thereby improving the precision of LLR. Based on simulations of the Yunnan Observatories' 1.2 m laser ranging system using the Apollo 11 CCR array, the echoes for 10 ps laser ranging are obtained when the reflector array is tilted 7° in the longitude direction while remaining untilted in the latitude direction. The tilt status of the reflector array inferred from the peaks and morphology of the echoes closely matches the simulated preset values. The single-shot ranging precision of this ranging is calculated to be 32.24 mm, while selecting the nearest peak as the ranging signal yields a single-shot ranging precision of 2.52 mm, significantly improving the single-shot ranging precision. Furthermore, the distance measurement obtained from each peak can be reduced to the center of the reflector, thereby improving the accuracy of the ranging data. Subsequent ground target experiments verify that as the tilt angle of the CCR array increases, the data width of ranging echoes and the resulting random error also increase, leading to a decrease in ranging precision. Analyses of ranging experiment echoes from arrays with two or three columns of CCRs validate the conclusion that the tilt angle of the CCR array can be inferred from the echoes of the waves, providing a basis for intercepting partial signals for ranging data processing, which is consistent with the theoretical analysis. When processing the echo signals from the three-column CCR array, higher-precision ranging data are obtained by selecting the middle peak for data processing. When performing laser ranging with an inclined CCR array, capturing the middle peak as the echo signal can also reduce the data advancement caused by the detector, thereby improving the accuracy of the laser ranging results. When performing laser ranging on a tilted array of CCRs, extracting the middle peak as the echo signal can reduce the data shift caused by the detector, thereby improving the accuracy of laser ranging results.

## 5 Discussion

Among the stations currently capable of performing LLR tasks, the Geodetic Observatory Wettzell is now using a 10 ps laser to conduct LLR work. Figure 20 shows the simulated laser echo envelope (Figure 20a), the simulated echo signal histogram (Figure 20b), the actual observed data for the Apollo 15 CCR array from this station (Figure 20c), and the histogram of the measured data (Figure 20d). The actual measurement data used in the figure were collected on 2 March 2023 from 18:53:20 to 22:18:47 UTC and represent full−rate data aggregated from 22 segments of ranging measurements of the Apollo 15 CCR array on that day. The data source and the parameters used in the simulation system for this station were both obtained from the ILRS. The average lunar libration for the corresponding time period at this station was calculated to be $-3.97°$ in latitude and $2.93°$ in longitude. Considering that the Apollo-series CCR arrays had an initial placement deviation of about $±1°$ when deployed on the lunar surface, the simulation used a lunar libration of $-2.97°$ in latitude and $1.33°$ in longitude. As shown in the figures, the echo in this ranging experiment at Geodetic Observatory Wettzell exhibited a multi−peak pattern similar to the theoretical echo, but due to the limited number of data points, it cannot fully reflect the tilted state of the CCR array. With advancements in laser technology, particularly the further enhancement of laser pulse energy, and the adoption of a 10 ps high−repetition−rate laser ranging system, it will be possible to obtain echo signals from CCR arrays at different tilt angles. By extracting data from multiple peaks of the echo signals, not only can the precision of LLR be improved, but the tilt state of the CCR arrays can also be studied based on the echo signals. This allows the ranging results to be attributed to the center of the CCR array, thereby increasing the accuracy of LLR and advancing the development of the Earth–Moon model towards higher precision.

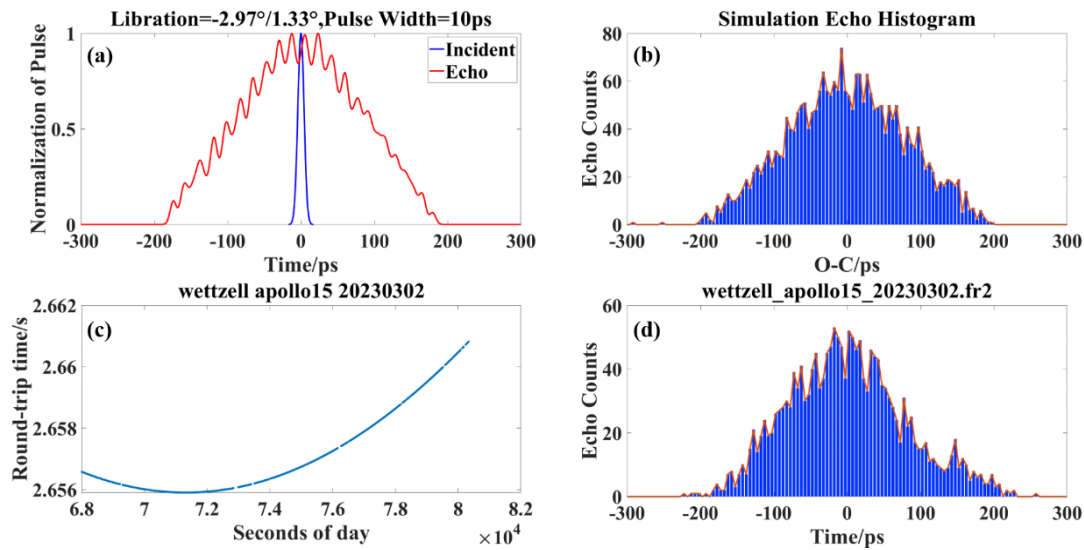

Figure 20 The experimental results for the CCR array with three columns.

## References：

[1]　BENDER P L, CURRIE D G, POULTNEY S K. The Lunar Laser Ranging Experiment[J/OL]. Science, 1973, 182(4109): 229-238. DOI:10.1126/science.182.4109.229.

[2]　DICKEY J O, BENDER P L, FALLER J E, Lunar Laser Ranging: A Continuing Legacy of the Apollo Program[J/OL]. Science, 1994, 265(5171): 482-490. DOI:10.1126/science.265.5171.482.

[3]　MURPHY JR T W. Lunar laser ranging: the millimeter challenge[J/OL]. Reports on Progress in Physics, 2013, 76(7): 076901. DOI:10.1088/0034-4885/76/7/076901.

[4]　CHABÉ J, COURDE C, TORRE J, Recent Progress in Lunar Laser Ranging at Grasse Laser Ranging Station[J/OL]. Earth and Space Science, 2020, 7(3): e2019EA000785. DOI:10.1029/2019EA000785.

[5]　COURDE C, TORRE J M, SAMAIN E, Lunar laser ranging in infrared at the Grasse laser station[J/OL]. Astronomy & Astrophysics, 2017, 602: A90. DOI:10.1051/0004-6361/201628590.

[6]　BATTAT J B R, MURPHY T W, ADELBERGER E G, The Apache Point Observatory Lunar Laser-ranging Operation (APOLLO): Two Years of Millimeter-Precision Measurements of the Earth-Moon Range1[J/OL]. Publications of the Astronomical Society of the Pacific, 2009, 121(875): 29-40. DOI:10.1086/596748.

[7]　MURPHY T W, ADELBERGER E G, STRASBURG J D, Testing Gravity via Next-Generation Lunar Laser-Ranging[J/OL]. Nuclear Physics B - Proceedings Supplements, 2004, 134: 155-162. DOI:10.1016/j.nuclphysbps.2004.08.025.

[8]　MERKOWITZ S M. Tests of Gravity Using Lunar Laser Ranging[J/OL]. Living Reviews in Relativity, 2010, 13(1): 7. DOI:10.12942/lrr-2010-7.

[9]　WILLIAMS J G, BOGGS D H, YODER C F, Lunar rotational dissipation in solid body and molten core[J/OL]. Journal of Geophysical Research: Planets, 2001, 106(E11): 27933-27968. DOI:10.1029/2000JE001396.

[10]　WILLIAMS J G, NEWHALL X, DICKEY J O. Lunar moments, tides, orientation, and coordinate frames[J/OL]. Planetary and Space Science, 1996, 44(10): 1077-1080. DOI:10.1016/0032-0633(95)00154-9.

[11]　TURYSHEV S G, WILLIAMS J G, FOLKNER W M, Corner-cube retro-reflector instrument for advanced lunar laser ranging[J/OL]. Experimental Astronomy, 2013, 36(1-2): 105-135. DOI:10.1007/s10686-012-9324-z.

[12] SAMAIN E, MANGIN J F, VEILLET C, Millimetric Lunar Laser Ranging at OCA (Observatoire de la Côte d'Azur)[J/OL]. Astronomy and Astrophysics Supplement Series, 1998, 130(2): 235-244. DOI:10.1051/aas:1998227.

[13] MURPHY T W, ADELBERGER E G, BATTAT J B R, The Apache Point Observatory Lunar Laser-ranging Operation: Instrument Description and First Detections[J/OL]. Publications of the Astronomical Society of the Pacific, 2008, 120(863): 20-37. DOI:10.1086/526428.

[14] MURPHY T, STRASBURG J, STUBBS C, APOLLO: Meeting the Millimeter Goal[R]. In Proc. 14th International Workshop on Laser Ranging. 2004., 2004.

[15] MARTINI M, DELL'AGNELLO S, CURRIE D, MoonLIGHT: A USA–Italy lunar laser ranging retroreflector array for the 21st century[J/OL]. Planetary and Space Science, 2012, 74(1): 276-282. DOI:10.1016/j.pss.2012.09.006.

[16] CURRIE D, DELL'AGNELLO S, DELLE MONACHE G. A Lunar Laser Ranging Retroreflector Array for the 21st Century[J/OL]. Acta Astronautica, 2011, 68(7): 667-680. DOI:10.1016/j.actaastro.2010.09.001.

[17] DEHANT V, PARK R, DIRKX D, Survey of Capabilities and Applications of Accurate Clocks: Directions for Planetary Science[J/OL]. Space Science Reviews, 2017, 212(3): 1433-1451. DOI:10.1007/s11214-017-0424-y.

[18] ZHANG M, MÜLLER J, BISKUPEK L, Characteristics of differential lunar laser ranging[J/OL]. Astronomy & Astrophysics, 2022, 659: A148. DOI:10.1051/0004-6361/202142841.

[19] KOKURIN Y L. Lunar laser ranging: 40 years of research[J/OL]. Quantum Electronics, 2003, 33(1): 45-47. DOI:10.1070/QE2003v033n01ABEH002363.

[20] ALLEY C O, CHANG R F, CURRI D G, Apollo 11 Laser Ranging Retro-Reflector: Initial Measurements from the McDonald Observatory[J/OL]. Science, 1970, 167(3917): 368-370. DOI:10.1126/science.167.3917.368.

[21] FALLER J E, BENDER P L, ALLEY C O, Geodesy Results Obtainable with Lunar Retroreflectors[M/OL]//The Use of Artificial Satellites for Geodesy. American Geophysical Union (AGU), 1972: 261-264[2024-05-22].

[22] ĆUK M, HAMILTON D P, LOCK S J, Tidal evolution of the Moon from a high-obliquity, high-angular-momentum Earth[J/OL]. Nature, 2016, 539(7629): 402-406. DOI:10.1038/nature19846.

[23] HUANG K, ZHANG L, YANG Y, Dynamical Model of Rotation and Orbital Coupling for Deimos[J/OL]. Remote Sensing, 2024, 16(7): 1174. DOI:10.3390/rs16071174.

[24] WILLIAMS J G, BOGGS D H, TURYSHEV S G, Lunar Laser Ranging Science[M/OL]. arXiv, 2004[2023-11-03]. http://arxiv.org/abs/gr-qc/0411095.

[25] NEWHALL X X, WILLIAMS J G. Estimation of the Lunar Physical Librations[J/OL]. International Astronomical Union Colloquium, 1997, 165: 21-30. DOI:10.1017/S0252921100046339.

[26] YANG Y Z, LI J L, PING J S, Determination of the free lunar libration modes from ephemeris DE430[J/OL]. Research in Astronomy and Astrophysics, 2017, 17(12): 127. DOI:10.1088/1674-4527/17/12/127.

[27] SCHEFFER L K. Better Lunar Ranges with Fewer Photons - Resolving the Lunar Retro-reflectors[M/OL]. arXiv, 2005[2023-10-12]. http://arxiv.org/abs/gr-qc/0504009.

[28] DEGNAN J J. Millimeter accuracy satellite laser ranging: A review[M/OL]//SMITH D E, TURCOTTE D L. Geodynamics Series: 25. Washington, D. C.: American Geophysical Union, 1993: 133-162[2023-10-18]. DOI:10.1029/GD025p0133.